\documentclass[manuscript]{aastex}
\usepackage{natbib}

\shorttitle{Site testing of the Sierra de Javalambre}
\shortauthors{Moles et al.}

\begin{document}
%
\title{Site testing of the Sierra de Javalambre. First results}
\author{
M.~Moles\altaffilmark{1,2},
S.F~S\'anchez\altaffilmark{1,3,4}, 
J.~L.~Lamadrid\altaffilmark{1},
A.~J.~Cenarro\altaffilmark{1},
D.~Crist\'obal-Hornillos\altaffilmark{1, 2},
N.~Maicas\altaffilmark{1},
J.~Aceituno\altaffilmark{4}
}

\altaffiltext{1}{Centro de Estudios de F\'\i sica del Cosmos de Aragon (CEFCA), C/General  Pizarro 1, E-41001 Teruel, Spain}
\altaffiltext{2}{Instituto de Astrof\'\i sica de Andaluc\'\i a (IAA), Consejo   Superior de Investigaciones Cient\'\i ficas (CSIC), C/ Camino Bajo de Hu\'etor 50, E-18008, Granada, Spain}
\altaffiltext{3}{Fundaci\'on Agencia Aragonesa para la Investigaci\'on y el Desarrollo (ARAID)}
\altaffiltext{4}{Centro Astron\'omico Hispano-Alem\'an, Calar Alto, (CSIC-MPG),   C/Jes\'us Durb\'an Rem\'on 2-2, E-04004 Almeria, Spain }

\email{moles@cefca.es}

\begin{abstract}
   
We present the main characteristics of the {\sl Pico del Buitre}, at the {\it
  Sierra de Javalambre}, the proposed location for the Javalambre
Astrophysical Observatory. The measurements have been obtained from
spectrophotometric, photometric and seeing data obtained with different
monitors and instruments on the site. We have also used publicly accessible
meteorological satellite data to determine the total time useful for
observations.

The night-sky optical spectrum observed in a moonless night shows very little
contamination by the typical pollution lines. Their contribution to the
sky-brightness is $\sim$0.06 mag in B, $\sim$0.09 mag in V and $\sim$0.06 mag
in R. In particular, the comparison of the strengths of the Sodium artificial
and natural lines indicates that the site satisfies the IAU recommendations
for a dark site. The zenith-corrected values of the moonless night-sky surface
brightness are B = 22.8 mag arcsec$^{-2}$, V = 22.1 mag arcsec$^{-2}$, R =
21.5 mag arcsec$^{-2}$, I = 20.4 mag arcsec$^{-2}$, which indicates that the
site is very dark. The extinction has been measured for the summer period,
with a typical value of 0.22 mag in the V-Band, with the best measured value
of 0.18 mag in a totally photometric night. The median value of the seeing in
the $V$-band for the last two years (2008-9) is $0.71\arcsec$, with a mode of
0.58$\arcsec$. The seeing values present a seasonal pattern, being smaller in
summer ($\sim$0.69$\arcsec$) than in winter time (0.77$\arcsec$). For 68\% of
the analyzed nights the seeing was better than 0.8$\arcsec$ during the entire
night. The seeing is found to be stable for rather long periods, in particular
for the nights with good seeing values. The typical scale, for nights with the
seeing below 0.8$\arcsec$, is about 5 hours for variations within 20\% of the
reference value. The fraction of totally clear nights is $\sim$53\%, while the
fraction of nights with at least a 30\% of the night clear is $\sim$74\%.

 \end{abstract}

%


\section{Introduction}

Any search for adequate sites for new observatories has to focus on the
observing conditions such as the night sky brightness, the number of clear
nights, the seeing, the transparency or the photometric stability. Looking for
a potential new site in continental Spain we found in 1989 that the {\sl
  Sierra de Javalambre}, in the province of Teruel, might be one such site. A
limited number of studies and observations were accumulated at that time
indicating that it was a promising site. For a variety of reasons the study
had to be interrupted until recently when the idea of building a new
Observatory there regained interest and new studies were funded. Consequently
we have started a program to determine the main observing conditions at the
proposed site for the Javalambre Astrophysical Observatory (JAO,
hereafter). The selected site is on the {\sl Pico del Buitre}, Sierra de
Javalambre (Teruel, Spain), 1957 m above the sea level, at
40$^o$02$\arcmin$28.67$\arcsec$ North, 01$^o$00$\arcmin$59.10$\arcsec$ West.

The JAO will house two new telescopes to be built in the next three years. It
will be run by the {\sl Centro de Estudios de F\'\i sica del Cosmos de
  Arag\'on}, CEFCA, recently created in Teruel. The main telescope is intended
to be a large {\sl Etendue} instrument for large scales surveys, with an
aperture of 2.5m and a FoV of 7 square degrees, with good quality over the
whole field. The first Astronomical Survey is being defined to match the
requirements to accurately determine the Baryonic Acoustic Oscillations along
the line of sight using photometric redshifts for Luminous Red Galaxies, as
proposed by the {\sl Physics of the Accelerated Universe}, PAU, project. The
Javalambre-PAU Astrophysical Survey will be a large scale ($\sim$8000
degree$^2$) photometric Survey with narrow band ($\sim$100~\AA) filters
covering the spectral range from $\sim$4500-8500~\AA. The details of the
project, its aims and implementation are given in (Ben\'\i tez et al. 2009a).

A smaller, auxiliary telescope of 80 cm of aperture will be also built. The
main task of this telescope will be to perform the required photometric
calibrations for the main survey. In addition, $\sim$30\% of the observing
time will be accessible to the astronomical community. The telescope will have
a FoV of $\sim$2 square degrees.

In this work we study the main characteristics of the night-sky at the {\sl
  Pico del Buitre}, identified as the best location in the Sierra de
Javalambre. The study is mostly focused on the last two
years (2008-09), which correspond to a period of rising solar activity from
the last
minimum\footnote{http://www.ngdc.noaa.gov/stp/SOLAR/ftpsolarradio.html}. The
derived main properties have been compared with similar properties at other
observatories. In Section 2 we describe the dataset collected for the current
study, including a description of the data reduction process; in Section 3 we
show the analysis performed on the different types of data and the results
derived for each one. The main results are summarized in the Section 4.


\section{Description and Analysis of the Data}
\label{data}

Data from different sources were collected to characterize the different aspects of the night-sky emission at the proposed location for
the JAO. These sources and the kind of data we used are described in the following subsections.

\subsection{Spectroscopic data}
\label{spec_data}

A 40~cm aperture telescope temporarily installed at the observatory, together
with a commercial DSS-7 spectrograph and a ST-8XME camera from
SBIG \footnote{http://www.sbig.com/} have been used to take night-sky
spectra. The main characteristics of the spectrograph are listed in Table
\ref{tab_spec}.

The observations were done the night of the 23rd of August 2009, three days
after the new-moon, under clear conditions. A series of exposures of 600
seconds of integration time each were taken along the night, with the
telescope pointing to the Zenith and without tracking. With this setup the
traces of possible astronomical sources across the slit can be easily
rejected. Moreover, only the spectra without contamination by the Milky Way
were considered for the analysis. The final number of spectra amounts to 14,
corresponding to a total integration time of 8400 seconds.

The data were reduced using the standard procedures included in IRAF and R3D
(S\'anchez 2006). First, the bias was determined for each frame using the non
illuminated areas of the chip and dark exposures taken at the beginning of the
night with the same exposure time. The bias frame was subtracted from the
corresponding science frame, which was subsequently trimmed to consider only
the illuminated region. A series of dark and twilight frames were used to
identify bad pixels: hot, dead and low-sensitivity ones. The values in those
pixels were replaced by the median of the 5 adjacent ones along the
cross-dispersion axis, in the science frames. The twilight frames were used to
derive a normalization response frame, obtained dividing them by the median
spectrum (i. e., the median, for each column, of the values in the
corresponding rows). This frame was used to normalize the response of the
detector.

For each frame the median spectrum was extracted, corresponding to a total of
40 rows. It was wavelength calibrated using the most prominent night-sky
emission lines. Then the  14 frames were combined, adopting a cosmic-ray
rejection scheme, to produce a single combined frame, interpolated to a common
wavelength step using a simple 4-order polynomial function (eg., S\'anchez
2006). The accuracy of the wavelength calibration was $\sim$1.3~\AA,
acceptable given the wavelength resolution of the spectrum for the purposes of
this study. The wavelength calibrated spectrum was relative flux calibrated by
adopting the efficiency curve published by the manufacturer. Finally, it was
scaled to a typical night-sky surface brightness for an astronomical dark-site
 in the $V$-band.


\subsection{Sky-Brightness and Extinction data}
\label{ext_data}

The night-sky surface brightness and extinction were monitored using an
instrument built on purpose by J. Aceituno, the Extinction Camera and
Luminance Background Register, EXCALIBUR. It is a robotic extinction monitor
able to do quasi-simultaneous photometric observations in 8 bands covering the
wavelength range between 3400~\AA\ to 10200~\AA. The sky-brightness and
extinction curve are derived by comparing the instrumental and catalogue
photometry of pre-calibrated stars. About 16 extinction coefficients for each
of the sampled bands are estimated every hour, together with up to $\sim$50
extinction coefficients per band and per night. The same kind of instrument
has been used to study the extinction curve at the Calar Alto observatory
(S\'anchez et al. 2007).

The instrument was operative at the mountain for a total of 13 nights, between
the 23rd of July and the 5th of October 2008. For those nights, it was setup
to obtain the sky-brightness at $B$ and $R$ bands, and the extinction at
$\sim$500~nm, which basically corresponds to the $V$-band. A total of 118
estimations of the night sky surface brightness for the broad-bands, and 317
accurate estimations of the extinction for the narrow-band filters, were
derived.

\subsection{Seeing data}
\label{seeing_data}

The night-sky seeing has been measured with a robotic Differential Image
Motion Monitor (RoboDIMM, Aceituno 2004). This instrument measures the seeing
at the wavelength corresponding to the Johnson $V$-band. It was calibrated
with a similar unit installed at the Calar Alto observatory (Vern\'in \&
Mu\~noz-Tu\~noz 1995; S\'anchez et al. 2007) that was itself calibrated with a
Generalized Seeing Monitor in May 2002 \cite{ziad05}. The values measured by
this instrument were also checked with simultaneous direct imaging obtained
with the 3.5m Telescope at Calar Alto \cite{sanc08}.

The instrument was first installed at the {\sl Pico del Buitre} in March 2008,
and it has been systematically used since then, provided that the weather
conditions and the still incipient infrastructure could permit it. The data
presented here include measurements up to the end of September 2009. Apart
from adverse meteorological conditions, the coverage is not complete due to
several kinds of accidents that affected the infrastructure.

RoboDIMM measures the transversal and horizontal components. Occasionally
there is a large difference between these two values, most probably due to
mechanical oscillations, that have to be eliminated from the
statistics. Thanks to the inherent stability of the instrument the frequency
of such departures is rather low, less than $\sim$5\% in our case. The number
of collected measurements amounts to 81651 in 132 nights. Despite the
technical difficulties we encountered to work on the mountain, this coverage
is similar to that of another well established observatories (eg., S\'anchez
et al. 2007).

\subsection{Satellite Information}

We used data from the satellites METEOSAT (148 images) and LANDSAT (2175
images) corresponding to the period 1983-1988 during the early site testing of
the Sierra de Javalambre. They represent a faithful first approximation to the
analysis of the cloud coverage in the area. Both sets of images were
cross-checked and used to find the global properties of the Sierra and the
differences between different peaks.

More recently, publicly available high resolution images corresponding to the
period 2005-2006 were analyzed. The data-base consist on 3,000 images from the
satellite NOAA (sampling of 1.2 km per pixel) and 2,000 images taken with the
remote sensors MODIS-TERRA and MODIS-AQUA (sampling of 250 m per pixel). The
images correspond to different moments of the day and night, UT = 0h, 06h, 12h
and 18h. Terra MODIS and Aqua MODIS are viewing the entire Earth's surface
every 1 to 2 days, acquiring data in 36 spectral bands covering the wavelength
range between 0.4~$\mu$m to 14.4~$\mu$m. These images can be used to determine
the presence of clouds at a given location, on the basis of the difference of
temperature between the surface and the clouds, provided by the infrared
measurements. The analysis was done on images of 20$\times$20 km centered on
the {\sl Pico del Buitre}.

\section{Results and Discussions}
\label{ana}

We describe here the analysis performed over each of the different sets of data. 

\subsection{Night Sky Spectrum}
\label{ana_spec}

Figure 1 shows the night-sky spectrum obtained as explained
before. This is the first time that a spectrum of the night-sky over
Javalambre is published. The night-sky emission lines have been labeled with
its name and wavelength. For comparison purposes, published night-sky spectra
corresponding to the Calar Alto and the Kitt Peak observatories have been
included in the figure.


The main natural contribution to the night-sky light is airglow. It produces
prominent lines like OI$\lambda\lambda$5577, 6300~\AA, the OH bands in the red
and NIR and a pseudo-continuum in the blue (due to overlapping O$_2$ bands,
2600-3800~\AA), and in the green (NO$_2$ bands, 5000-6000~\AA). It also
contributes to the ubiquitous NaD$\lambda$5890/6~\AA~ doublet, a line that can
be contaminated by light-pollution from low and high pressure street-lamps
(see Figure 1). A more detailed description of the effects of the
airglow on the night-sky emission can be found in Benn \& Ellison (1998a). An
atlas of the airglow from 3100~\AA~ to 10000~\AA~ was presented by Ingham
(1962) and Broadfoot \& Kendell (1968). Other contributions to the night-sky
spectrum are the zodiacal light, the starlight, and the extragalactic light,
which increase the background continuum emission (see Benn \& Ellison 1998a
and references therein).

 Apart form the natural contributions, the night sky spectrum can be affected
by light-pollution, mostly from street-lights in populated areas near
to observatories. Light-pollution arises principally from tropospheric
scattering of light emitted by Sodium (high and low pressure), Mercury-vapor
and incandescent street lamps (McNally 1994; Osterbrock et al., 1976; Holmes
1997). The Na low pressure lamps are those with less impact on astronomical
observations, since it produces most of its light concentrated in the
5890/6~\AA~ and 8183/95~\AA~ lines. High-pressure sodium lamps emit most of
its light in a broad, FWHM$\sim$400~\AA~ Na line centered at $\sim$5890~\AA~
that shows a central reversal. They also show strong emission at 8183/95~\AA~
and fainter emission lines. The Mercury lamps produce narrow lines at
3651/63~\AA, 4047~\AA, 4358~\AA, 5461~\AA, 5770~\AA~ and 5791~\AA, together
with broad features at 6200~\AA~ and 7200~\AA, FWHM$\sim$100~\AA, from the
phosphor used to convert UV to visible light. They also produce a weak
continuum emission over the whole visible range.

The spectrum of the incandescent lamps consists of continuum emission only,
and it is difficult to identify in a night-spectrum. Finally, the
high-pressure metal halide lamps, nowadays frequently used in the illumination
of sport stadiums and monuments, exhibit some Scandium, Titanium and Lithium
emission lines, that are characterized by a blue edge due to molecular bands
(General Electric 1975; Lane \& Garrison 1978; Osterbrock et al. 1976).

The night-sky spectrum of the {\sl Pico del Buitre} illustrated in
Figure 1 shows little evidence of significant light pollution from
street-lamps. The Mercury lines, that are present in the night-sky spectrum of
the Calar Alto and Kitt-Peak observatories, are not detected. Moreover, the
broad emission at $\sim$5900~\AA~ from high-pressure sodium lamps is
significantly weaker than in these two observatories, whereas the potentially
strong emission line at 5893~\AA~ from low-pressure sodium lamps is barely
detected at most. The detected contamination, that includes the Sc line at
5351.1~\AA, is dominated by the illumination of the city and metropolitan area
of Valencia ($\sim$810,000 habitants), at $\sim$100 km towards the South-East.
In this regards the night-sky spectrum is more similar to that of the
Observatory of the Roque de los Muchachos, at La Palma, published by Pedani
(2005).

To quantify the contribution of the light pollution to the night-sky spectrum
of the {\sl Pico del Buitre} we have determined the flux intensity
corresponding to each detected emission feature using the procedure described
in S\'anchez et al. (2007). The values we have found are listed in Table
\ref{tab_line}.  We also give in the Table the flux from the Sodium broad-band
emission, estimated by fitting the observed feature with a single broad
gaussian function. All the fluxes were converted to Rayleigh following the
conversion formulae by Benn \& Ellison (1998a).

Quantifying the artificial contribution to the Sodium lines is difficult
because of the natural contribution to them. From the analysis by Benn \&
Ellison (1998a) this natural contribution to the broad Sodium emission band
can be estimated to be $\sim$0.04 mag in the V and R bands. Adopting this
value, the contribution from the artificial component can be easily
estimated. The results are listed in Table~\ref{tab_cont} for the $B$, $V$ and
$R$ bands, and for a set of medium band filters used by the ALHAMBRA survey
(Moles et al., 2008; Ben\'\i tez et al., 2009b). They are included to
illustrate the effects of the pollution lines when median/narrow-band filters,
as proposed for large-scale surveys (see Ben\'\i tez et al., 2009b), are
used. The results show that the contribution to the natural component at the
Javalambre site in all bands is small. In fact, the site of {\sl Pico del
  Buitre} satisfies the IAU criterium for a dark site since the artificial
contribution to the Sodium emission, estimated to be $\leq 0.06$ mag, doesn't
exceed that of the natural airglow (Smith 1979).

A long term monitoring of the night-sky spectrum at the observatory will be
required to analyze the evolution of the light pollution along the time. Most
major observatories nearby heavily populated areas have some kind of night-sky
protection laws. Although the Sierra de Javalambre site does not benefit yet
from any local sky-protection law to regulate the street illumination, it
appears that its effect is low in the night-sky spectrum. In any case a proper
protection law would be most convenient to preserve the quality of the site.


\subsection{Night Sky Brightness and Atmospheric Extinction}
\label{ana_mag}

The data from EXCALIBUR were used to determine the night sky brightness in the
$B$ and $R$ bands and the extinction in the $V$ band. The sky brightness
measurements were not corrected for extinction, following the convention
adopted in most of the recent studies of sky brightness (e.g., Walker 1988b;
Krisciunas 1990; Lockwood et al 1990; Leinert et al. 1995; Mattila et
al. 1996; Benn \& Ellison 1998a,b, S\'anchez et al. 2007).  They are only
corrected to the zenith using the expression by Patat (2003),

$$\Delta m = -2.5 {\rm log}_{10}[(1-f)+f X]+\kappa (X-1)$$

\noindent
where $\Delta m$ is the increase in sky brightness at a given band for an
airmass $X$, $f$ is the fraction of the total sky brightness generated by
airglow, being (1-$f$) the fraction produced outside the atmosphere (hence
including zodiacal light, faint stars and galaxies) and $\kappa$ is the
extinction coefficient at the corresponding wavelength. To apply that formula
we have used the average extinction along each night and a typical value of $f
=$ 0.6 (Patat 2003; S\'anchez et al. 2007). The collected data spans two
complete moon phases, having a good coverage of the different
phases/illumination of the moon.


Table~\ref{tab_mag} lists the mean values of the sky brightness for dark nights, i. e., with Moon illumination less than 5\%. It contains the measured $B$ and $R$ values and the estimated values for the $V$ and $I$ bands. The later were obtained from the night-sky spectrum scaled to the $B$ and $R$ values measured for dark nights. In the same table values for other observatories are given. It appears that Javalambre is one of the darkest observatories even if it has to be taken into account that some of the data in the Table (for example for Paranal) were obtained during the maximum of solar activity.  


The measurements of the extinction obtained for Javalambre correspond to the
summer season when higher than year average extinction values are
expected. This is due to the presence of aerosols and, very occasionally at
the Javalambre latitude, to dust from the Sahara desert. The overall effect is
clearly seen in the seasonal evolution of the dust extinction at well
established observatories in Spain, like La Palma (Benn \& Ellison 1998a) or
Calar Alto (S\'anchez et al. 2007).

Considering all the 317 individual measurements of the extinction coefficient
provided by EXCALIBUR the median value amounts to 0.27 mag, with a high
dispersion of 0.14 mag. Inspecting the data set we verify that 4 nights have
very high extinction coefficients, over 0.4 mag. These were highly
non-photometric nights with a very unusual dust content in the
atmosphere. Excluding them, the median value for k$_V$ is now 0.22, much
closer to the expected values and very similar to that of the summer period in
Calar Alto. Indeed, truly photometric nights are characterized by rather small
extinction variations along the night. For the best night in our sampling,
October 5, 2008, the variations of k$_V$ are within 15\% and the night
extinction coefficient is $\sim$0.17$\pm$0.03 mag, a value consistent with
that reported for the winter season in other major observatories, when the
aerosol contribution is lower (Benn \& Ellison 1998b; S\'anchez et al. 2007).
Despite of that, the average extinction values reported so far seem to be
rather high compared with the statistical values reported in other
observatories. A visual inspection of the extinction curves derived by
EXCALIBUR for different wavelengths for the different nights (eg., S\'anchez
et al. 2007), points to an excess of Ozone absorption, rather than a high
aerosol content in the vicinity of the observatory. The limited amount of data
collected so far does not allow us to perform a more detail analysis, that
will require a larger dataset. For similar reasons, no seasonal pattern in the
extinction was analyzed due to the short time coverage of the collected data.



\subsection{Atmospheric Seeing}
\label{ana_seeing}


The seeing data described in section \ref{seeing_data} were used to determine an average seeing for each monitored night along the  $\sim$1.5 years covered by the data discussed here. In Figure~\ref{seeing1} the nightly measured data points are shown. The obvious lack of data for some periods is due to technical problems related to the rather precarious situation we still have at the mountain to get the monitors working and safe from accidents. In fact the two longer periods without data do correspond to two major events: damage by a strong wind and a robbery. Looking at the nightly average values (Figure~\ref{seeing1}), we see that $\sim$68\% of the nights have median seeing below 0.8$\arcsec$. Considering the whole data set of all the validated measurements, amounting to N = 81651 data points, the median value we find is 0.71$\arcsec$. The distribution is shown in Figure~\ref{seeing2}. It can be seen that the mode is 0.58$\arcsec$ and more than 85\% of the values are below 1$\arcsec$. 

The trend of the average nightly values that can be appreciated in
Figure~\ref{seeing1} is suggestive of a seasonal pattern, as it is found in
other places (S\'anchez et al. 2007). To further investigate this possibility
we have built two subsets of data, one for May-September (Summer period) and
the other for October-April (Winter season). With some caution due to the lack
of the data for January and February, its is clear from Figure~\ref{seeing2}
the seeing for the Summer period (median value of 0.69$\arcsec$) is better
than for the Winter period (median 0.77$\arcsec$). Not only the median seeing
is better in the Summer season, but the chances of having better seeing in a
Summer night are much higher.

In Table \ref{tab_seeing} we have collected the median seeing determined for Javalambre and published data for other observatories. It appears that the median seeing at Javalambre is comparable to that of the most highly reputed observatories in the world.

We have looked for the possible dependence of the seeing value with some
meteorological parameters. We have found strong indication of correlation only
with the wind speed and direction. We have selected all the seeing
measurements that are coincident within 5 minutes with measurements of the
wind direction, a total of 16542 data points. We find that the seeing worsens
for wind speeds over 18 m/s, a completely expected result. More interesting is
the relation between seeing quality and wind direction. Figure~\ref{seeing3}
shows the median seeing for different wind directions. It is clear that the
dominant winds are from western directions and the best seeing values are
found under these wind conditions (median = 0.60$\arcsec$). The worst seeing
values are found when the wind comes from the North or North-East directions
($\sim1\arcsec$). Fortunately these are the less frequent wind directions.

We have also considered the stability of the seeing. Inspecting the data we
find that the seeing can be very stable. Indeed, it is frequent to find
nights, particularly under good seeing conditions ($< 0.8\arcsec$)
presenting seeing variations of just a few per-cent. To characterize the
stability of the seeing on an objective basis we have worked out the
typical time scales to have the same seeing within some percentage of the
reference value. For that we have analyzed the whole data set computing the
point-to-point variations within a given percentage. Then the distribution of
the resulting time scales for every percentage were computed. The result is
plotted in Figure~\ref{seeing_st}. We find that the time scale for the seeing
stability is longer for better seeing values. It can be as long as 5 or more
hours for variations up to 20\% for reference seeing values below
0.8$\arcsec$.

\subsection{Fraction of Useful Time}
\label{ana_sat}

The analysis of the low resolution images from METEOSAT and LANDSAT for the
period 1983-1988 showed that the fraction of completely clear days was about
53\%. They also showed that the SE side of the Sierra de Javalambre has the
best statistics so all the efforts were subsequently concentrated on the {\sl
  Pico del Buitre}, the highest peak in that specific area.

The analysis of the high resolution images corresponding to the period
2005-2006 confirms the earlier values, indicating that no long-term evolution
in the overall conditions has occurred. It is confirmed that the number of
totally clear nights per year amounts to 193 or $\sim$53\%. Another important
result is that the number of nights with less than 50\% cloud cover
amounts to 62.2\% (227 nights per year). We notice that these figures compare
very well with highly reputed Observatories around the world (see for example,
Webster 2004; S\'anchez et al. 2007).

The fraction of useful time has a clear seasonal pattern, being higher in
Summer than in Winter. We show in Figure~\ref{weather} the fraction of time
with cloud coverage lower than a certain limit (10\%, 50\% and 70\% of the
time, for each night), for the different months along a year. The time period
between June and August has the better weather statistics, with $\sim$60\% of
totally clear nights. On the other hand, the worst months are March, May and
November, with $\sim$35\% of clear nights. These months are traditionally the
periods with the largest precipitation values in continental Spain.

\section{Conclusions}
\label{conc}

We have characterized the main properties of the night-sky at the {\sl Pico del Buitre}, Sierra de Javalambre (Teruel, Spain), the site proposed for the Javalambre Astrophysical Observatory. Data taken with seeing, extinction and night-sky brightness {\sl in situ} monitors along a period of 1.5 years, together with satellite high spatial resolution meteorological data for the years 2005 and 2006 has been used to determined the relevant parameters for optical observations. The main conclusions that have been reached are:
 
\begin{itemize}

\item A night-sky spectrum, covering the wavelength range form 3950~\AA~ to 8150~\AA, for the moonless dark-time at the observatory has been presented for the first time. Airglow and light-pollution emission lines are detected in this spectrum. The strength of the light-pollution lines has been measured, estimating their contribution to the emission in different bands. The light pollution is weak or absent (for the Mercury lines, for example). The {\sl Pico del Buitre} fulfills the IAU recommendations for a dark astronomical site (Smith 1979).

\item The moonless night-sky brightness at the zenith has been measured in the $B$ and $R$ bands, and deduced for the $V$ and $I$-bands. It is found that the site is particularly dark with marginal contribution at most by artificial pollution.  
 
\item The extinction, estimated for the Summer season, shows a typical value of $k_V =$ 0.22 mag, similar to that of other observatories for the same season. The extinction found for a photometric night was $k_V\sim$0.17 mag, consistent with the
expected value for a night without aerosols.

\item The median seeing along the last 1.5 years was 0.71$\arcsec$, with 68\% of the nights with $< 0.8\arcsec$ median value. The seeing shows a seasonal dependence being better in the Summer period. These values put the {\sl Pico del Buitre} among the first rank known observatories. 

The seeing shows a clear dependency with the wind direction. Only for N and NE directions the seeing increases to near 1$\arcsec$. For all the other directions of the wind rose the typical value is $\sim$0.60$\arcsec$. The  frequency of the N-NE wind direction is less than 10\%.

The seeing at the observatory is remarkably  stable. Our measurements indicate that in good seeing conditions ($< 0.8\arcsec$), the seeing can be stable for more than 5 hours within 20\% of the reference value.

\item The fraction of completely clear (less than 10\% cloud coverage) nights
  amounts to 53\%. The fraction of useful nights (at least a 30\% of the night
  clear), amounts to 74\%. So far, the fraction of photometric nights
  is still not known for the observatory, with the current data.

\end{itemize}

We conclude that the {\sl Pico del Buitre}, the proposed site to build the Javalambre Astrophysical Observatory is a particularly good astronomical site. The fact that Javalambre doesn't present special accessing difficulties and that it is placed in continental Europe represents an added advantage from the point of view of building and operation. 

\section{Acknowledgments}

The {\sl Ministerio de Ciencia e Innovaci\'on}, the {\sl Consejo Superior de
  Investigaciones Cient\'\i ficas} and the {\sl Departamento de Ciencia,
  Innovaci\'on y Universidad} of the {\sl Gobierno de Arag\'on} are
acknowledged for support to the site testing work and the JAO project.

SFS thanks the Spanish Plan Nacional de Astronom\'\i a program
AYA2005-09413-C02-02, of the Spanish Ministery of Education and Science and
the Plan Andaluz de Investigaci\'on of Junta de Andaluc\'{\i}a as research
group FQM322.
  
We thanks the Calar Alto Observatory, and in particular his director,
Dr. J. Alves, to allow us to use their spare DIMM for the first 6 months of
2008 to start this study. We also thanks the {\sl Instituto Tecnol\'ogico de
  Arag\'on} for help with the management of the site testing and technical
advice.

We thank Dr. J. Mu\~noz, Dr. M. A. Torrej\'on, J. A. Quesada and S. Abad for
their help during the early studies at {\sl Pico del Buitre}. We also thank
E. Esco for his support and involvement at the beginning of the new period of
measurements.  We also thank the members of the Amateur Astronomer Society in
Teruel ({\it ACTUEL}), F.~Garc\'\i a, M.~Mart\'\i n, J.A.~Ort\'\i z and
J.~Bull\'on, for the help with some of the observations.

We also thanks for Dr.~J.~Melnick anf Dr.~M.~Sarazon for their advice in the
preparation of this study.

\newpage

%
%

\begin{table}
\begin{center}
\caption{Main properties of the DDS-7 Spectrograph}
\label{tab_spec}
\begin{tabular}{llrr}\hline
\tableline\tableline
Parameter   & Value \\
\tableline
Slit-Width         &  $\sim$2.7$\arcsec$\\
Slit-Length         &  $\sim$19$\arcsec$\\
Spectral Resolution (FWHM)&  15 \AA \\
Spectral Sampling  & 5.4 \AA/pixel \\
Spectral Range     & 4000-8100 \AA \\
Peak Efficiency    & 0.4 \\
Wavelength of the Peak Efficiency & 6000~\AA \\
\tableline
\end{tabular}
\end{center}
\end{table}

\begin{table}
\begin{center}
\caption{Properties of the detected emission lines}
\label{tab_line}
\begin{tabular}{llrr}\hline
\tableline\tableline
Line Id & Wavelength & Flux$^*$ & Flux$^{**}$\\
        & (\AA)      &   & R \\
\tableline
HgI     &4827,32 & 0.4$\pm$0.1 &  2.9\\
NaI     &4978,83 & 1.2$\pm$0.1 &  8.6\\
NaI     &5149,53 & 0.4$\pm$0.1 &  2.9\\
ScI     &5351.1  & 0.7$\pm$0.1 &  5.0\\
HgI     &5461    & 0.9$\pm$0.1 &  6.5\\
OI      &5577    & 6.8$\pm$0.1 & 48.8\\
NaI     &5683,88 & 1.2$\pm$0.1 &  8.6\\
HgI     &5770,91 & 0.5$\pm$0.1 &  3.6\\
Broad NaD&5893   &  6.3$\pm$0.2 & 45.2\\
NaI     &6154,61 & 0.8$\pm$0.1 &  5.7\\
OI      &6300    & 2.1$\pm$0.1 & 15.1\\
OI      &6364    & 0.9$\pm$0.1 &  6.5\\
\tableline
\end{tabular}

$*$ in units of 10$^{-16}$ erg~s$^{-1}$~cm$^{-2}$\\

$**$ in units of rayleighs.
\end{center}
\end{table}

\begin{table}
\begin{center}
\caption{Contribution of the light pollution lines to the sky-brightness}
\label{tab_cont}
\begin{tabular}{ll}\hline
\tableline\tableline
Band  & $\Delta$mag\\
\tableline
$B$   & 0.06 \\
$V$   & 0.09 \\
$R$   & 0.06 \\
4280/331 & 0.04\\
5510/331 & 0.05\\
5820/331 & 0.22\\
\tableline
\tableline
\end{tabular}
\end{center}
\end{table}

\begin{table*}
\begin{center}
\caption{Summary of the night-sky surface brightness}
\label{tab_mag}
\begin{tabular}{llllll}\hline
\tableline\tableline
Site \& Date &         B       &    V     &    R    &   I   & Reference\\  
\tableline
Javalambre   &     22.8$\pm$0.6 &  22.1$\pm$0.5 &  21.5$\pm$0.3 &  20.4$\pm$0.5  & This work\\
\tableline\tableline
La Silla      1978 &   22.8     &    21.7     &  20.8       & 19.5  & Mattila et al. (1996)\\
Kitt Peak     1987 &   22.9     &    21.9     &             &       & Pilachowski et al. (1989)\\
Cerro Tololo  1987 &   22.7     &    21.8     &  20.9       & 19.9  & Walker (1987a,88a)\\
Calar Alto 1990    &   22.6     &    21.5     &  20.6       & 18.7  & Leinert et al. (1995)\\   
La Palma 1990-92   &   22.5     &    21.5     &             &       & Benn \& Ellison (1998a,b)\\
La Palma 1994-96   &   22.7     &    21.9     &  21.0       & 20.0  & Benn \& Ellison (1998a,b)\\
Mauna Kea 1995-06  &   22.8     &    21.9     &             &       & Krisciunas (1997)\\
Paranal  2000-01   &   22.6     &    21.6     &  20.9       & 19.7  & Patat et al. (2003)\\
MtGraham 2000-01   &   22.86    &    21.72    &  21.19      &       & Taylor et al. (2004)\\ 
Cerro Pach\'on 2005  &   22.43    &    21.63    &             & 20.3  & Walker \& Schwarz (2007)\\
Calar Alto 2007    &   22.86    &    22.01    &  21.36      & 19.25 & S\'anchez et al. (2007) \\
MtGraham 2008      &   22.81    &    21.81    &  20.82      & 19.78 & Pedani (2009) \\
\tableline
\end{tabular}

$*$ Based on spectrophotometric data obtained with PPAK.

\end{center}
\end{table*}

%
\begin{table}
\begin{center}
\caption{Median seeing in the $V$-band compared with other astronomical sites}
\label{tab_seeing}
\begin{tabular}{lrl}\hline
\tableline\tableline
Site & Median seeing & Reference \\
\tableline
Javalambre          & 0.71$\arcsec$\ \ \  & This work \\
Javalambre (Winter) & 0.77$\arcsec$\ \ \  & ``     `` \\
Javalambre (Summer) & 0.69$\arcsec$\ \ \  & ``     `` \\
\tableline
Mauna Kea (1987)    & 0.50$\arcsec$\ \ \  & Racine (1989)\\
Paranal (1993)      & 0.64$\arcsec$\ \ \  & Murtagh \& Sarazin (1993)\\
Paranal (2002-2007)  & 0.65$\arcsec$\ \ \ &  Sarazin et al. (2008) \\ 
La Palma (1997)      & 0.76$\arcsec$\ \ \   & Mu\~noz-Tu\~non et al. (1997)\\
La Silla  (1999)     & 0.79$\arcsec$\ \ \  & ESO webpage$^{*}$\\
Paranal (2005)       & 0.80$\arcsec$\ \ \  & ESO webpage$^{**}$\\
La Silla (1993)      & 0.87$\arcsec$\ \ \  & Murtagh \& Sarazin (1993)\\
Calar Alto (2006-2007)    & 0.90$\arcsec$\ \ \  & S\'anchez et al. (2007)\\
MtGraham (1999-2002) & $\sim$0.97$\arcsec$\ \ \  & Taylor et al. (2004)\\
Paranal (2006)       & $\sim$1.00$\arcsec$\ \ \  & ESO webpage$^{***}$\\
KPNO    (1999)       & $\sim$1.00$\arcsec$\ \ \ & Massey et al. (2000)\\
Lick (1990-1998)     & $\sim$1.90$\arcsec$\ \ \  & MtHamilton webpage$^{****}$\\
\tableline
\tableline
\multicolumn{3}{l}{}\\
\multicolumn{3}{l}{(*) http://www.ls.eso.org/lasilla/seeing/ }\\
\multicolumn{3}{l}{(**) http://www.eso.org/gen-fac/pubs/astclim/paranal/seeing/adaptive-optics/statfwhm.html}\\
\multicolumn{3}{l}{(***) http://www.eso.org/gen-fac/pubs/astclim/paranal/seeing/singstory.html}\\
\multicolumn{3}{l}{(****) https://mthamilton.ucolick.org/techdocs/MH\_weather/obstats/seeing.html}\\
\end{tabular}
\end{center}
\end{table}

  \begin{figure}
\resizebox{\hsize}{!}
{\includegraphics[width=\hsize,angle=270]{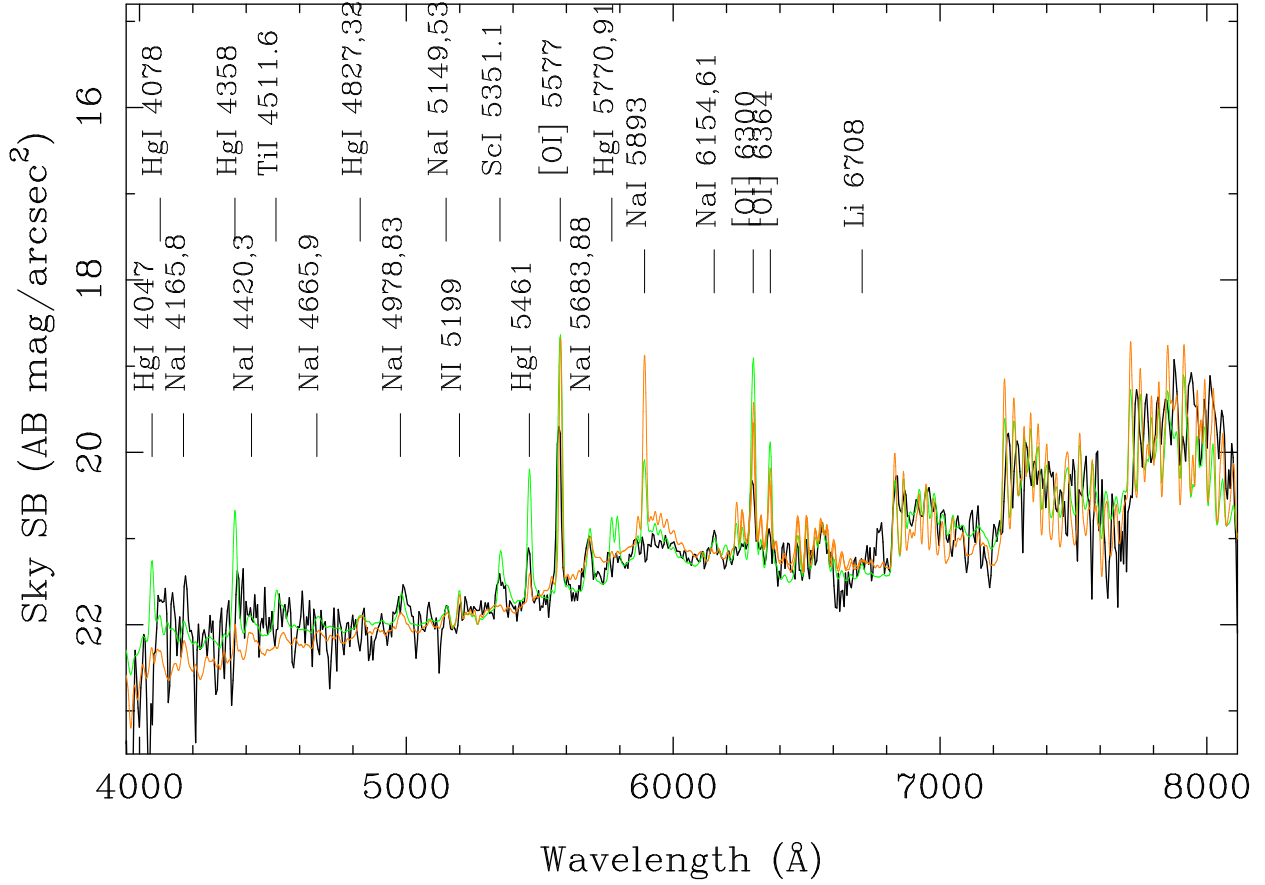}}
\label{spectrum}
  \caption{
Night-sky spectrum at {\sl Pico del Buitre}, Sierra de Javalambre, in the
optical wavelength range (3700-7950~AA), obtained the night of the 23rd of
August 2009, with dark-time and photometric conditions (black-line). The intensity has been
scaled to that of the typical moonless night sky-brightness in the
$V$-band. The most important night-sky emission lines have been labelled even
if they are not present in the spectrum. The NaI broad emission centered at
$\sim$5900~\AA, and the water vapor Meinel bands are clearly identified in the
spectrum. For comparison purposes we have included the night sky spectrum at
the Calar Alto Observatory (green line, S\'anchez et al. 2007) and that of the
Kitt Peak observatory (orange line, Massey \& Foltz 2000). The spectral
resolution and the sky-brightness at 5000~\AA, have been scaled to match our
data. The low intensity of the pollution lines at {\sl Pico del Buitre} is
clearly appreciated in this Figure (eg., the NaI~5893~\AA and Hg~5770,91~\AA lines).
 }
  \end{figure}

  \begin{figure}
\resizebox{\hsize}{!}
{\includegraphics[width=\hsize,angle=270]{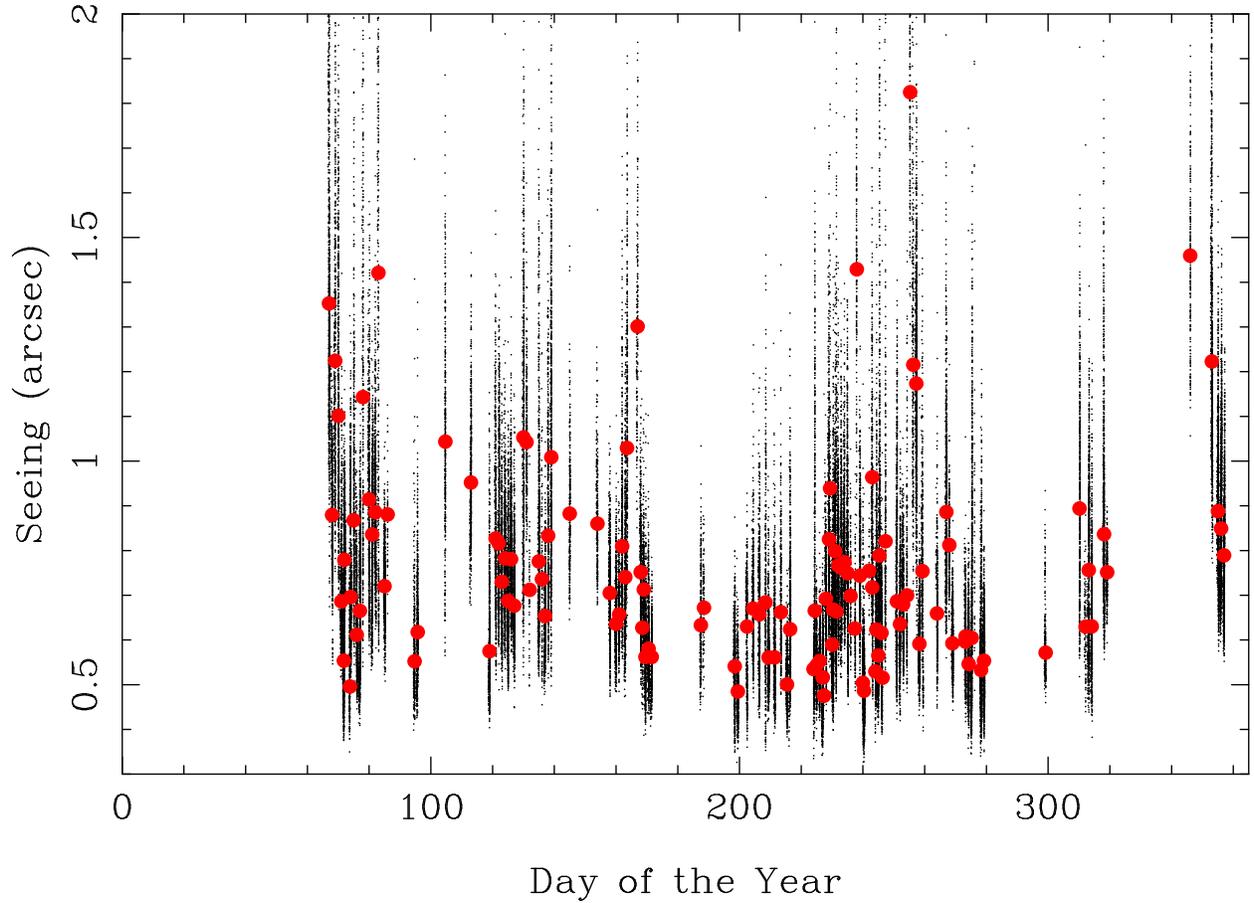}}
  \caption{\label{seeing1} Distribution of the seeing collected for the 132
    nights from March 2008 to September 2009 with RoboDIMM data, along the
   day of the year. The black-dots shows the 81651 individual data points, and
   the red-solid circles the median value for each night. It is appreciated a
   considerable coverage of the year, appart from January and February. 
 }
  \end{figure}

 \begin{figure}
\resizebox{\hsize}{!}
{\includegraphics[width=\hsize,angle=270]{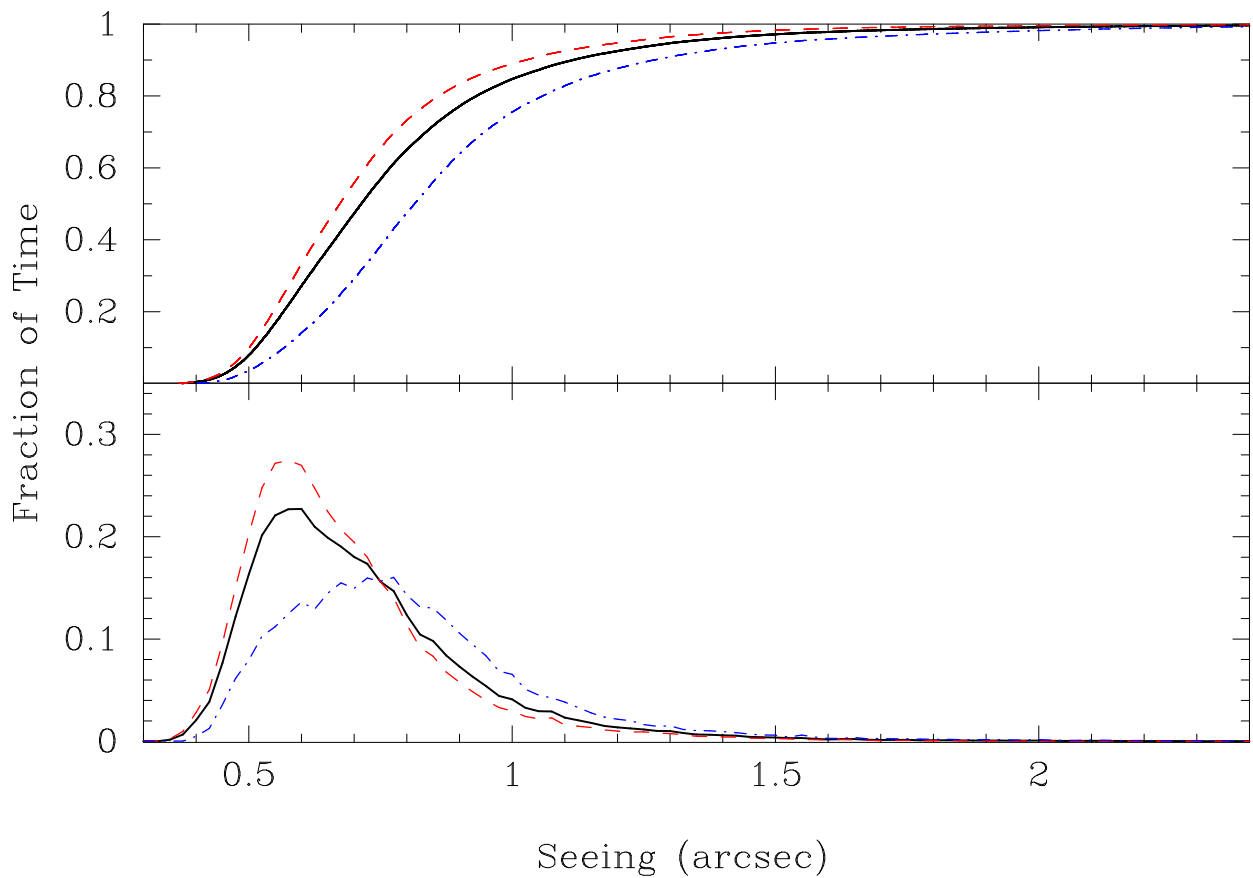}}
  \caption{\label{seeing2} 
Top panel: Cumulative distribution of the seeing values shown in Fig.~\ref{seeing1}, normalized to one (solid line), together with similar distributions for the seeing data corresponding to the Summer (red
dashed-line), and Winter (blue dashed-dotted line) seasons. Bottom panel: Normalized histogram of the cumulative distributions shown in the top panel (same symbols). The median value of the seeing is $\sim$0.71$\arcsec$, with $\sim$68\% of the nights with the median seeing below 0.8$\arcsec$. The distributions show that the seeing tends to be better in Summer than in Winter.
}
  \end{figure}

\begin{figure}
\resizebox{\hsize}{!}
{\includegraphics[width=\hsize,angle=270]{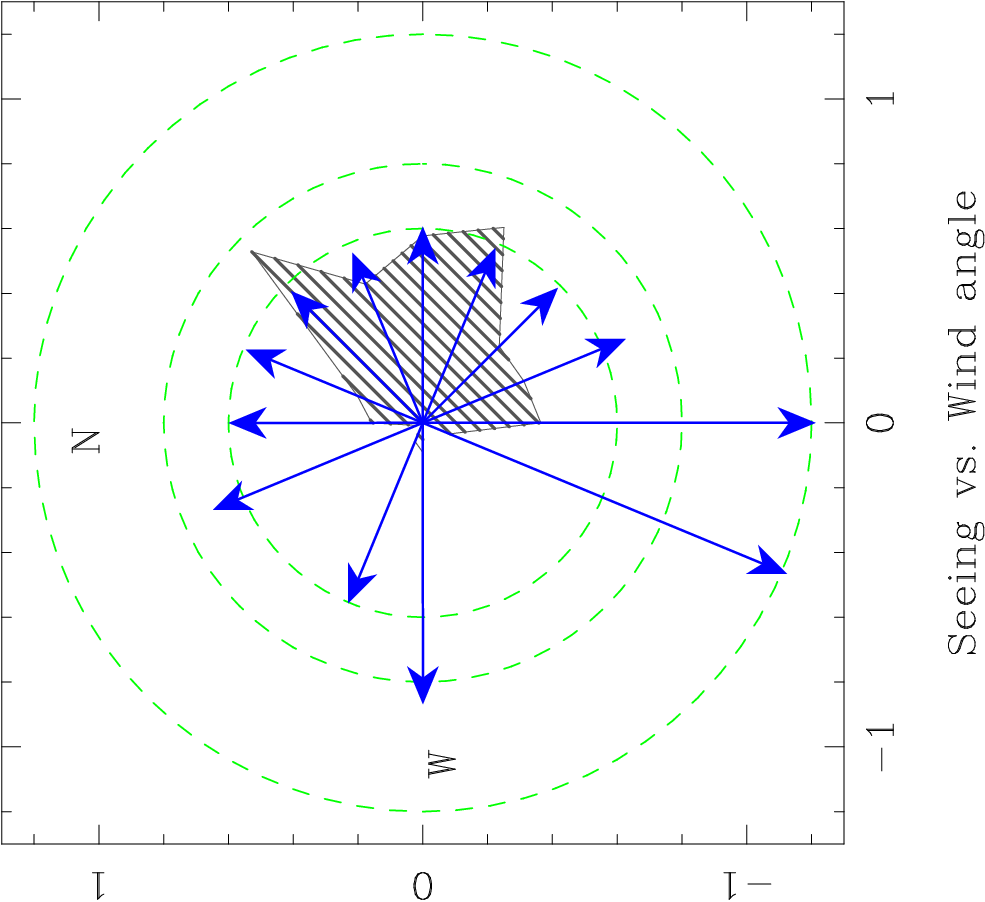}}
  \caption{\label{seeing3} 
Seeing distribution for different wind directions. Each arrow indicates the
wind direction and its length gives the corresponding typical seeing
value. The dashed-line circles indicates three different seeing values,
0.6$\arcsec$, 0.8$\arcsec$ and 1.2$\arcsec$. The polygonal line indicates the
relative frequency of each wind direction, normalized to one and scaled by a
factor 5 to make it more clearly visible.
}
  \end{figure}

 \begin{figure}
\resizebox{\hsize}{!}
{\includegraphics[width=\hsize,angle=270]{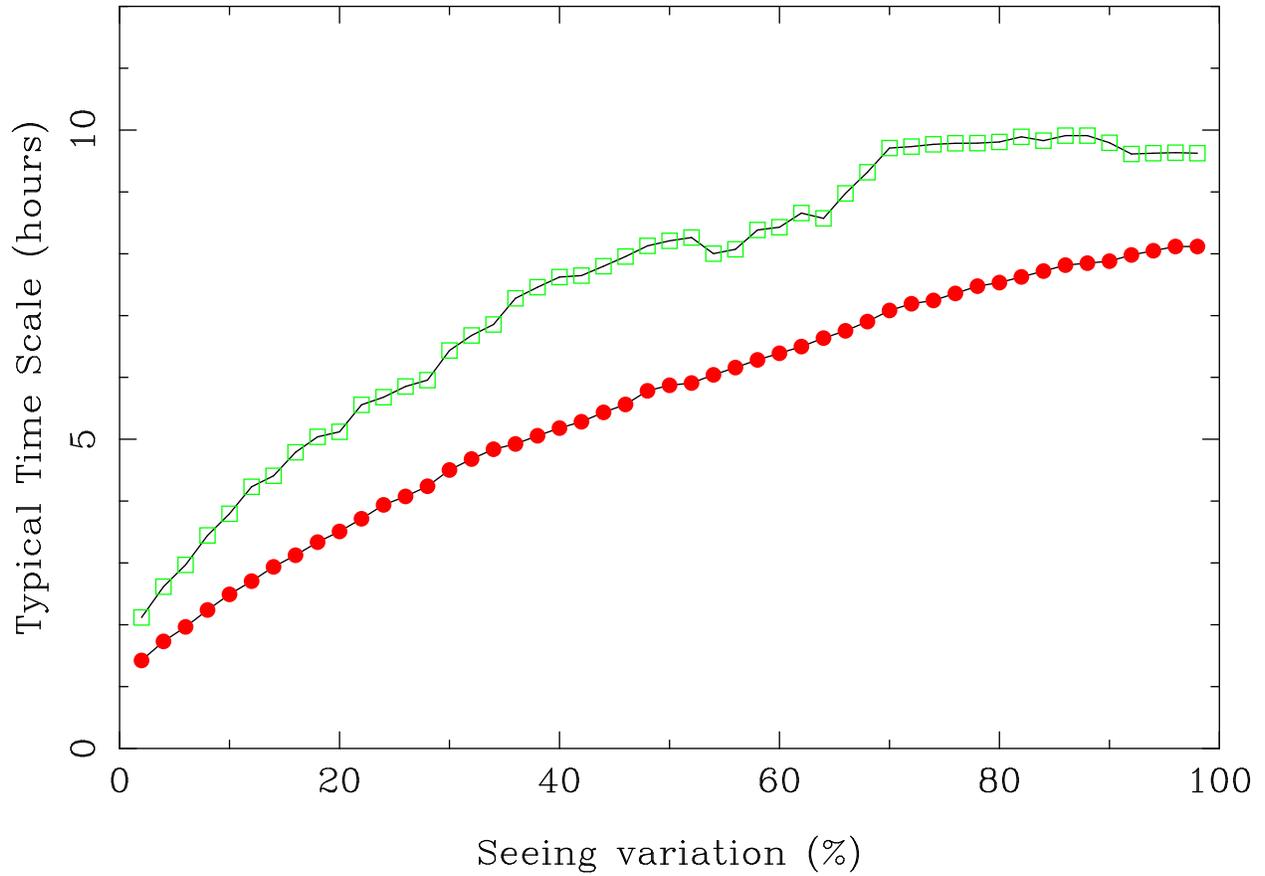}}
  \caption{\label{seeing_st} 
Seeing stability: Typical time scale for seeing variability within a given percentage of the reference value. The red solid circles are for any value of the reference seeing. The green symbols correspond to the cases when the reference seeing value is below 0.8$\arcsec$ 
}
  \end{figure}

 \begin{figure}
\resizebox{\hsize}{!}
{\includegraphics[width=\hsize,angle=270]{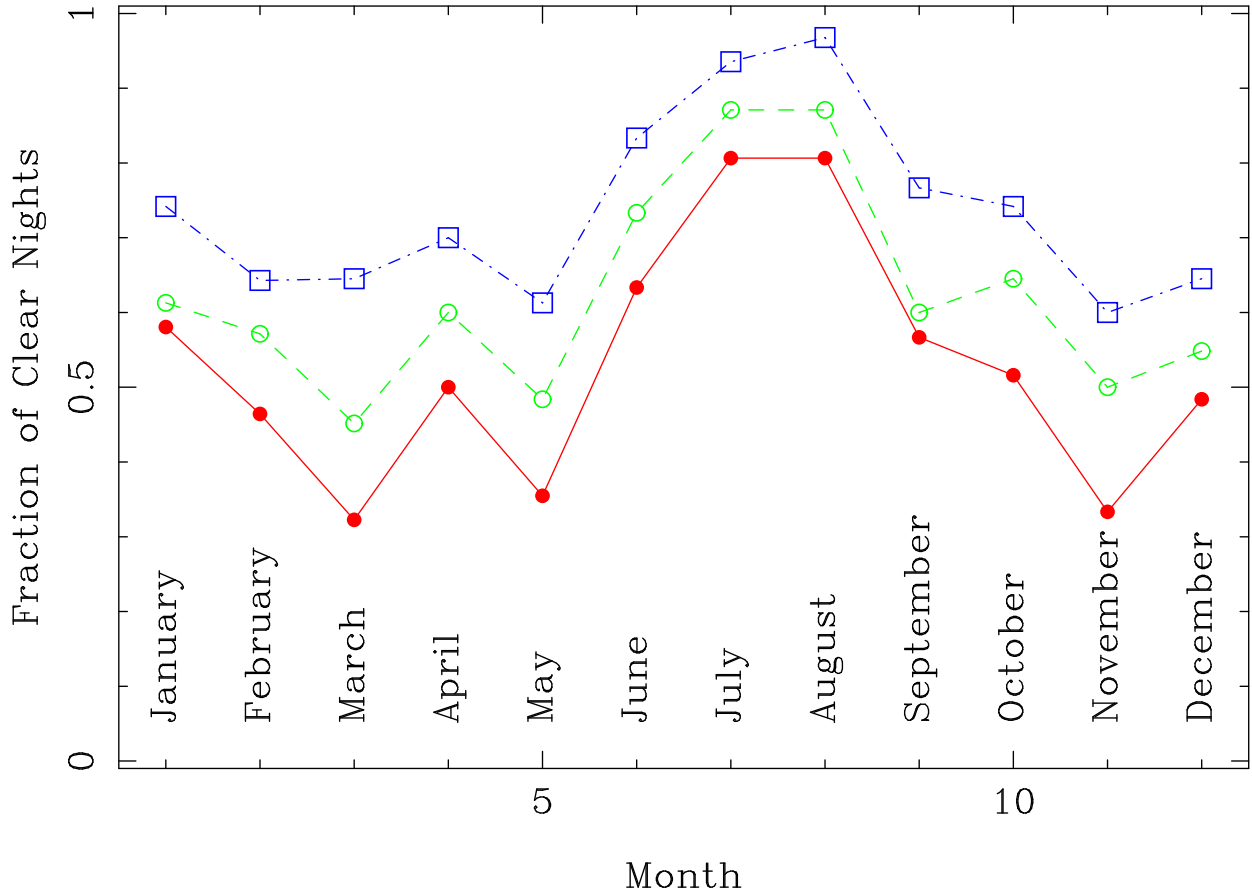}}
 \caption{\label{weather} 
 Clouds statistics along the different months for the time period 2005-2006, based on satellite images. The red-solid line and solid circles show the fraction of nights with a cloud coverage lower than 10\%. The green dashed-line and open circles show the fraction of nights with a cloud coverage lower than 50\%. Finally, the blue dashed-dotted line and open squares show the fraction of nights with a cloud coverage lower than a 70\%.
 }
  \end{figure}

 \end{document}